%% file: Island_mos2_arxiv_v2.tex
\documentclass[12pt]{iopart}
\usepackage{graphicx}
\usepackage{cite}
\begin{document}
\title{Thickness dependent interlayer transport in vertical MoS$_{2}$ Josephson junctions}
\author{Joshua O. Island$^1$, Gary A. Steele$^1$, Herre S.J. van der Zant$^1$, Andres Castellanos-Gomez$^2$}
\address{$^1$ Kavli Institute of Nanoscience, Delft University of Technology, Lorentzweg 1, 2628 CJ Delft, The Netherlands.}
\address{$^2$Instituto Madrileno de Estudios Avanzados en Nanociencia (IMDEA Nanociencia), Campus de Cantoblanco, E-28049 Madrid, Spain}
\ead{j.o.island@tudelft.nl}

\begin{abstract}
We report on observations of thickness dependent Josephson coupling and multiple Andreev reflections (MAR) in vertically stacked molybdenum disulfide (MoS$_{2}$) - molybdenum rhenium (MoRe) Josephson junctions. MoRe, a chemically inert superconductor, allows for oxide free fabrication of high transparency vertical MoS$_2$ devices. Single and bilayer MoS$_{2}$ junctions display relatively large critical currents (up to 2.5 $\mu$A) and the appearance of sub-gap structure given by MAR. In three and four layer thick devices we observe orders of magnitude lower critical currents (sub-nA) and reduced quasiparticle gaps due to proximitized MoS$_{2}$ layers in contact with MoRe. We anticipate that this device architecture could be easily extended to other 2D materials.       
\end{abstract}
\maketitle

\section{Introduction}
Transition-metal dichalcogenides (TMDC), and in particular MoS$_{2}$, have gained increased attention in the wake of the rise of graphene \cite{wang2012electronics, butler2013progress, buscema2015photocurrent}. In contrast with graphene, single-layer MoS$_2$ is a semiconductor with a sizable, direct band gap of 1.8 eV \cite{mak2010atomically}. Among the unique attributes of MoS$_2$, perhaps one of its most intriguing features is that the electronic band structure gradually changes with layer number\cite{mak2010atomically, splendiani2010emerging, eda2011photoluminescence, jin2013direct, lee2012mos2, lembke2015thickness}. Besides the recent attention on heterostructures\cite{britnell2012field, haigh2012cross, georgiou2013vertical, yu2013vertically, geim2013van}, transport measurements of the layer dependent properties of van der Waals materials have been limited to basal plane transport. Few works study c-axis (across layers) transport\cite{lee2011vertical, lee2013josephson, nguyen2014resonant, lee2015} and the dependence on flake thickness\cite{moriya2014large}. An interesting direction is the use of the well-known Josephson effect, whereby a supercurrent can flow between two superconductors connected by a tunnel barrier\cite{josephson1962possible, tinkham2012}, as a probe of the layer dependent electronic properties of MoS$_{2}$ flakes. The Josephson effect is not only uniquely sensitive to the type of weak link (metallic, insulating, and recent van der Waals junctions\cite{yabuki2016}) but also on the distance between two coupled superconductors. Recent theoretical works predict supercurrent reversal (0-$\pi$ transition) with back gate doping in planar, monolayer, MoS$_2$ Josephson junctions\cite{rameshti2014gate} but an experimental observation of Josephson coupling in MoS$_2$ junctions has not been reported. 

Here we probe the thickness dependence of interlayer electrical transport in MoS$_2$ flakes using the Josephson effect between two coupled molybdenum-rhenium (MoRe) superconductors. In addition to slow oxide growth of MoRe thin films\cite{seleznev2008deposition}, molybdenum itself has been shown to be an excellent contact metal for Schottky barrier-free contact to MoS$_{2}$\cite{kang2014, kang2014-2}. Using this alloy, we fabricate high transparency vertical junctions with 1-4 MoS$_{2}$ layers. In single and bilayer devices we observe high critical currents (up to 2.5 $\mu$A) and multiple Andreev reflections (MAR). In trilayer and four layer devices we observe orders of magnitude lower critical currents and the appearance of a reduced quasiparticle gap in the voltage carrying states. We attribute the metallic weak link behavior of the single and bilayer devices to strong hybridization between the MoRe contact and MoS$_{2}$ which leads to metallic MoS$_{2}$ layers\cite{kang2014, kang2014-2}. For thicker flakes, transport occurs through first the proximitized layers due to hybridization and then by tunneling through non-hybridized (uncoupled) MoS$_{2}$ layers.  

Fabrication of the vertical junctions is accomplished by first exfoliating commercially available MoS$_{2}$ onto a flexible polydimethylsiloxane (PDMS) substrate. Figure 1(a) shows a transmission mode optical image of an exfoliated flake on a PDMS stamp. The flake is then transferred onto a Si/SiO$_{2}$ substrate with prepatterned MoRe electrodes using an all dry viscoelastic stamping method \cite{castellanos2014}. The flake lies on top of the MoRe electrode and partially overlaps the SiO$_{2}$ substrate below (see Figure 1(b)). After transfer, top electrodes are patterned overlapping the flake and the bottom MoRe electrode. Figure 1(b) shows the final device after creating the top electrodes with e-beam lithography, sputtering of MoRe, and liftoff in warm acetone. The inset of Figure 1(b) shows a cartoon of the cross section of the device where MoRe electrodes sandwich layers of MoS$_{2}$. After fabrication, the thickness of the devices is verified using photoluminescence spectroscopy and atomic force microscopy (AFM) measurements on the portion of the flake on the SiO$_{2}$ substrate (see Supporting Information). We measured nine devices in detail and present transport measurements on junctions having 1-4 layers of MoS$_{2}$. All low temperature (30 mK - 1.2 K) measurements were performed using four terminal current bias configuration in a dilution fridge equipped with copper powder and RC filters. Figure 1(c) shows the room temperature characteristics of four devices after fabrication having 1 to 4 layers (henceforth named devices A through D, respectively). The measured voltage ($V$) is plotted as a function of current bias ($I_b$). In the inset we plot the zero bias resistance at room temperature (normalized to the area) as a function the number of layers. The data do not follow a simple exponential dependence as would be expected for pure tunneling through a barrier. The Josephson effect, being sensitive to the type of weak link (metallic or tunnel barrier), provides us with a clear explanation of the layer dependence below.  

\begin{figure}
\centerline{\includegraphics [width=6in]{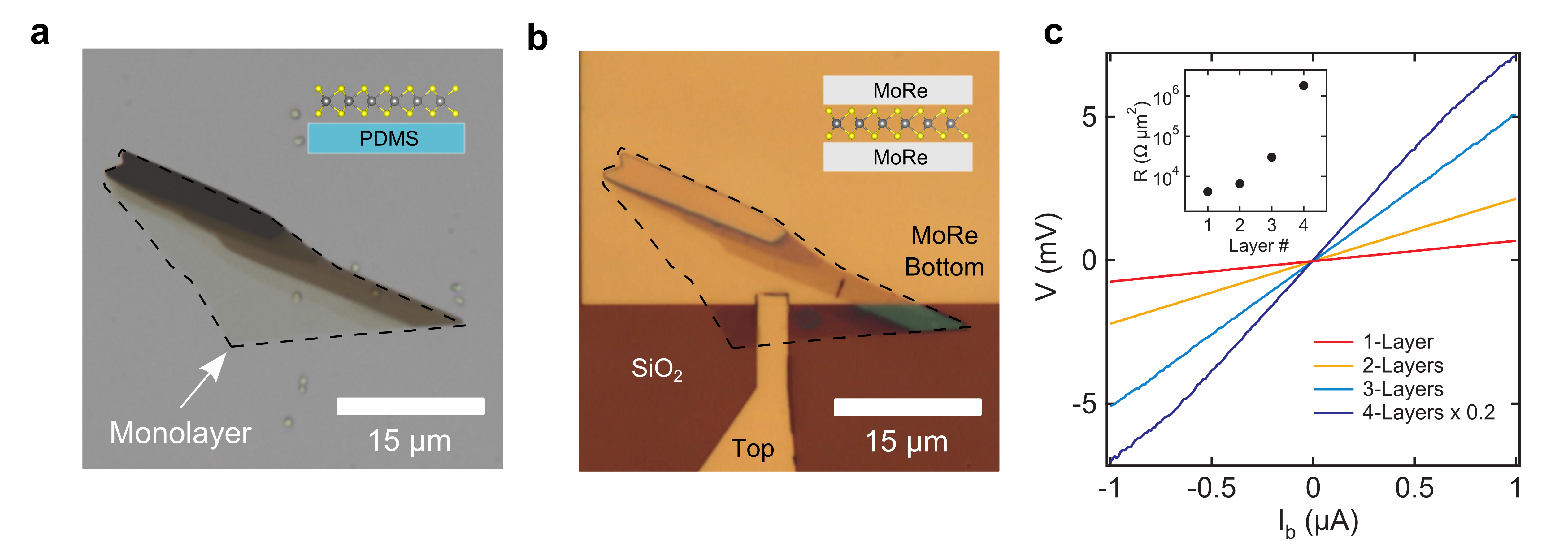}}
\textbf{Figure 1.} Fabrication of molybdenum-rhenium (MoRe) - molybdenum disulfide (MoS$_{2}$) vertical Josephson junctions (a) Transmission mode optical image of an exfoliated MoS$_{2}$ flake on a PDMS stamp. (b) Optical image of the same MoS$_{2}$ flake transferred onto a Si/SiO$_{2}$ substrate with a prepatterned MoRe bottom electrode. After transfer, a top electrode is patterned to sandwich the MoS$_{2}$ flake (see inset). (c) Room temperature voltage ($V$) as a function of current bias ($I_b$) for junctions with 1-4 layers of MoS$_{2}$ (devices A-D, respectively).
\end{figure}

\section{MoS$_{2}$ Josephson Junctions}
We now turn to the low temperature characteristics of the vertical MoRe-MoS$_{2}$ junctions. Figure 2 shows transport measurements for a junction with a monolayer MoS$_{2}$ flake (Device E) at 30 mK. A measurement of the voltage ($V$)  across the junction as a function of $I_b$ (Figure 2(a)) reveals a supercurrent that switches to the normal state at 2.5 $\mu A$. By irradiating the junction with an RF field, we observe the appearance of Shapiro steps signaling the coupling of the Josephson junction to the RF field \cite{shapiro1963, tinkham2012}. The inset of Figure 2(a) shows a color map of the differential resistance ($dV/dI$) as a function of $I_b$ (vertical axis) and RF power (horizontal axis). As the power is increased, more Shapiro steps enter the bias window. Additionally, we measure the junction response to an external magnetic field up to 12 T applied parallel to the MoS$_2$ basal plane (see Supporting Information). An overall decrease of the critical current is distinguished for fields parallel to the sandwich but the magnetic field dependence for all devices studied shows stochastic switching of the critical current. As the experiments are performed far above the first critical field of MoRe, the presences of pinned voritices in the disordered superconducting leads could be responsible for such strong switching behavior in the critical current. We further conjecture that flux focusing and corrugations in the surface texture of the MoRe electrodes further complicate the magnetic field response. 

\begin{figure}
\centerline{\includegraphics[width=5in]{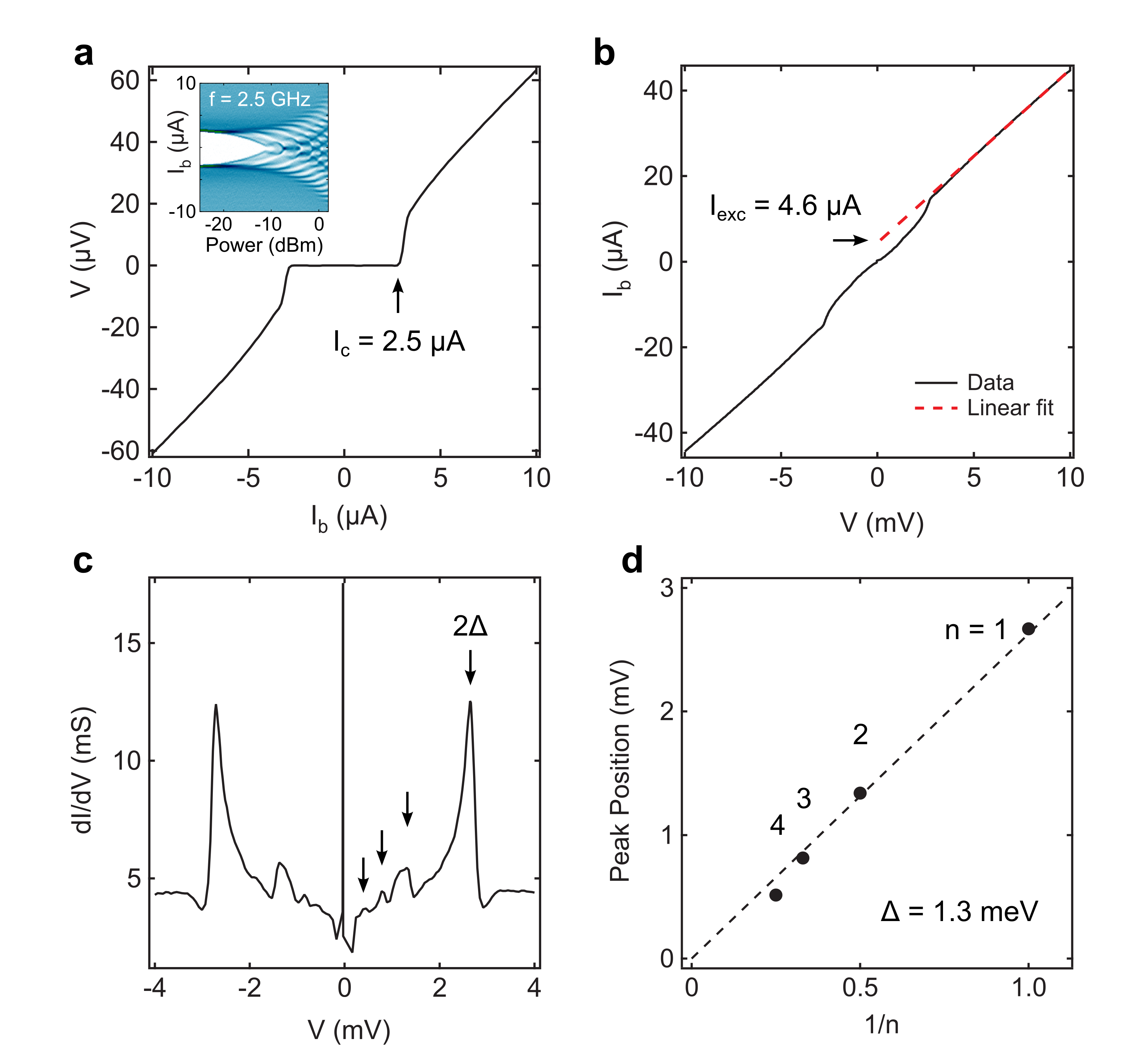}}
\textbf{Figure 2.} Single-layer MoS$_2$ Josephson junctions. (a) Plot of the measured voltage ($V$) as a function of current bias ($I_b$) for a single layer device (device E). Critical current of the junction is 2.5 $\mu$A. The inset shows the Shapiro steps  (dV/dI vs. $I_b$) as a function of RF power signaling the Josephson nature of the junction. The frequency of the irradiated field is 2.5 GHz. (b) $I_b$ vs. V for device A. The excess current ($I_{exc} = 4.6 \mu$A) is extrapolated from a fit (red dotted line) to the curve from 5 - 10 mV in the normal state. (c) Calculated differential conductance ($dI/dV$) as a function of $V$ for device A from data in panel (b). The arrows mark the conductance peaks arising from MAR. (d) The average peak position as a function of 1/$n$ for device A, where $n$ is an integer and corresponds to the number of Andreev reflections. The dotted line is a linear fit to the data which estimates a superconducting gap of $\Delta = 1.3$ meV. All measurements taken at 30 mK.
\end{figure}

Now turning to the voltage carrying state, in Figure 2(b) we plot $V$ vs. $I_b$ at higher bias currents for device A (at 30 mK). The curve becomes nonlinear below voltages of $\pm \approx 3$ mV. These nonlinear features are more clearly resolved in the differential conductance plotted in Figure 2(c) as a function of $V$. Symmetric peaks are observed at $\pm 2.6$ mV marking the onset of quasiparticle transport. Additionally, sub-gap conductance peaks are observed symmetric in voltage (see black arrows in Figure 2(c)). We attribute these sub-gap peaks to the well-known Andreev reflection process that takes place at the interface between a superconductor and a metal\cite{andreev1964thermal, klapwijk1982explanation}. The positions of the peaks in energy for multiple Andreev reflections (MAR) are given by $eV_n = 2\Delta /n$, where $n$ is a positive integer. In Figure 2(d) we plot the peak positions as a function of $1/n$. A linear fit to the average peak position results in a bulk superconducting gap estimate of $\Delta = 1.3$ meV. Assuming MoRe is a BCS superconductor ($\Delta = 1.76k_BT_c$)\cite{tinkham2012}, this estimate corresponds to a $T_c$ of 8.6 K which agrees closely with reported values for MoRe thin films \cite{witcomb1973, seleznev2008, singh2014}. MAR up to $n$ = 4 suggests relatively transparent transport barriers (transparencies between $\approx 10$\% and 100\%). We estimate the transparency of the interfaces given the excess current of the junction\cite{blonder1982transition}. The excess current is the extrapolated current at $V = 0$ V from a fit to the normal state current vs. voltage (see dotted line in Figure 2(b), here the fit is between 5 and 10 mV). For this single layer device we extract an excess current of 4.6 $\mu$A which corresponds to a contact transparency of $\approx$80\% \cite{blonder1982transition}. Similar high transparencies have been observed in Ge/Si\cite{xiang2006ge} and InAs\cite{doh2005tunable} Josephson junctions where Schottky barrier-free contact could be achieved.
       
\section{Thickness dependence}

In Figure 3 we present the low temperature layer dependence of the Josephson coupling for devices A-D having 1-4 layers. Figure 3(a-d) shows the $V-I_b$ curves for four devices measured at 1.2 K. The critical current decreases with increasing thickness but more importantly, we distinguish between relatively high critical currents for the single and bilayer devices and several orders of magnitude lower critical currents for the three and four layer junctions (see Supporting Information for critical current densities following the same trend and $I_cR_n$ products as a function of layer number). The voltage carrying state provides further insight on the layer dependence. In Figure 3(e-h) we plot $dI/dV$ for higher current biases measured at 30 mK for devices A-D. The dotted lines mark the bulk superconducting gap edge ($2\Delta = 2.6$ mV). While the single and bilayer junctions present high transparencies that give rise to MAR, the three and four layer junctions present opaque barriers resulting in the appearance of more well-defined quasiparticle gaps. We now discuss each regime in detail. 

Starting with the single layer device (Figures 3(e)), the sharp coherence peaks that align well with the bulk superconducting gap and the presence of MAR indicate a high transparency, metallic weak link. A possible explanation for this metallic behavior is doping of the MoS$_{2}$ flakes due to direct contact with the MoRe electrodes. Recently, Kang et al. have shown that molybdenum contacts provide tunnel barrier-free and Schotkky barrier-free contact to MoS$_{2}$ flakes \cite{kang2014}. In particular, DFT calculations show that due to the strong orbital overlap between Mo atoms in the electrode and the MoS$_{2}$ flake, the nearest single layer in top contacted devices becomes metallic \cite{kang2014, kang2014-2}. Subsequent layers below the top contacted layer remain semiconducting. A model of this scenario for the single layer device is shown in Figure 3(i) where the hybridized layer is colored blue. This is a reasonable explanation for the metallic behavior we observe which is supported below for the thicker devices as well. Additionally however, it should be noted that the presence of defects\cite{zhou2013, hong2015, addou2015} in the MoS$_{2}$ layers could give rise to metallic pinholes that would provide high transparency transport through the junction. A simple estimate for the size of a prospective pinhole can be made from the Sharvin resistance, the resistance of a metallic point contact: $R_s=4\rho l / 3\pi a^2$, where $\rho$ is the resistivity of the metal, $l$ is the mean free path of carriers, and $a$ is the radius of the pinhole\cite{sharvin1965, jansen1980}. If we assume the single layer devices are undoped semiconducting layers with pinholes, for our single layer devices (A and E), we estimate point contact radii of 1 nm and 2.5 nm, respectively (taking, for simplicity, a resistivity and mean free path of $\rho \approx 5$ $\Omega\mu$m and $l\approx 10$ nm for molybdenum\cite{yamashita1974}). These estimates are possible given the size of reported defects in exfoliated MoS$_2$\cite{addou2015}. Further investigation (junction statistics and cross sectional TEM) is required to determine the presence of pinholes in the high transparency junctions. 

The bilayer device (Figures 3(b, f, j)), also presenting high transparency ($\approx 80$ \%), indicates metallic weak link behavior. From the MAR conductance peaks (Figure 3(f)), following the analysis for device A in Figure 2(d), we estimate a reduced superconducting gap of $\Delta' = 1.1$ meV, slightly lower than the bulk value of $\Delta = 1.3$ meV from the single layer device. This indicates a thicker metallic link between the superconducting banks as compared with the single layer devices. Additionally, the magnetic field dependence shows better resolved oscillations of $I_c$ which point to more homogeneous current distributions (see Supporting Information). In the bilayer case, each layer is coupled to the nearest MoRe electrode providing transport through two hybridized MoS$_{2}$ layers (Figure 3(j)). 

\begin{figure}
\centerline{\includegraphics[width=6in]{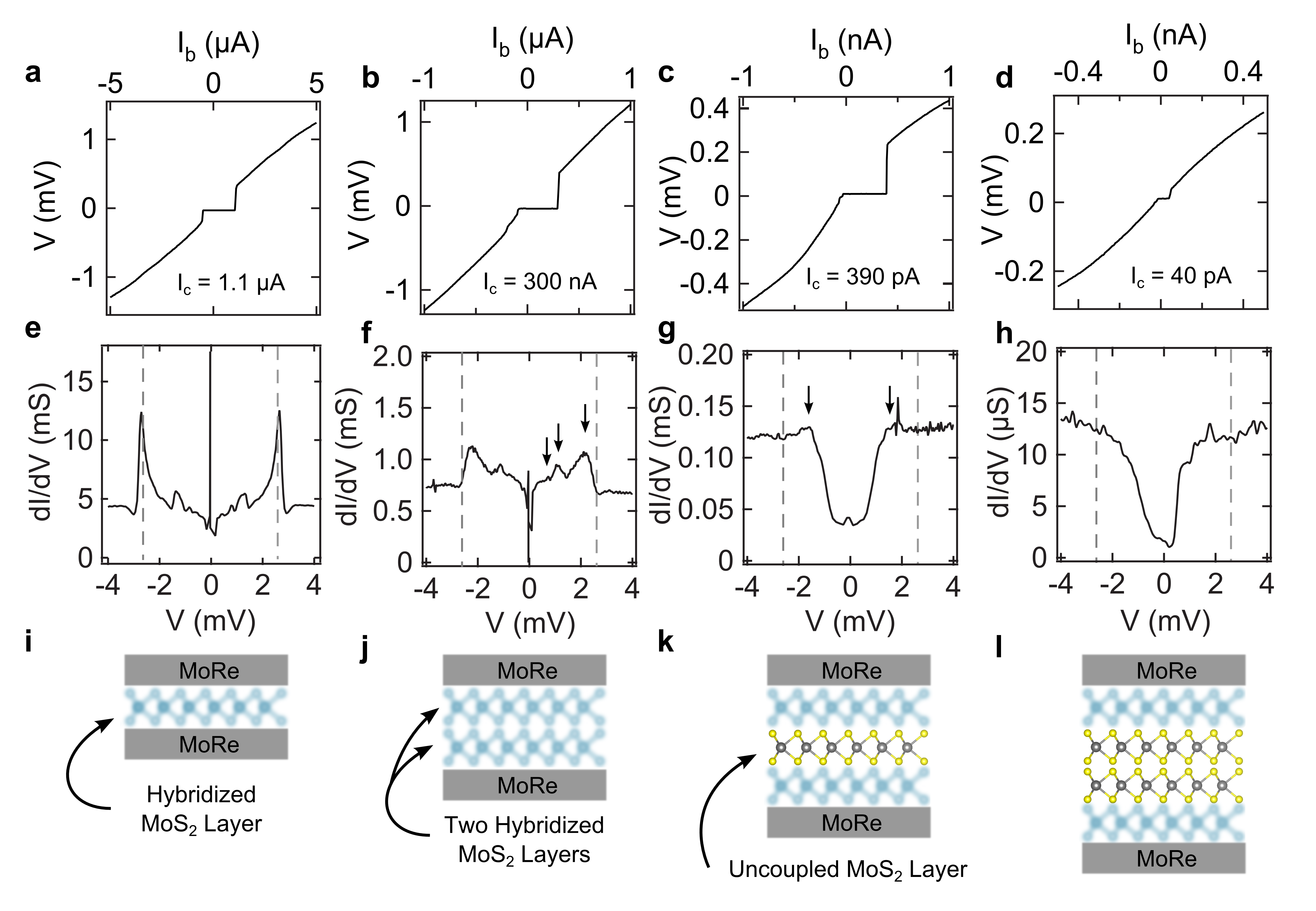}}
\textbf{Figure 3.} Thickness dependent Josephson coupling. (a-d) Current bias sweeps at 1.2 K for devices A-D having 1-4 layers of MoS$_2$, respectively. (e-h) Differential conductance ($dI/dV$) vs. V for devices A-D having 1-4 layers of MoS$_2$, respectively, taken at 30 mK. Panel (e) is the same data in Figure 2(c) plotted here for comparison. (i-l) Simple models explaining the layer dependence for each junction having 1-4 layers, respectively. The blurry blue MoS$_2$ layers represent the hybridization with the MoRe contacts suggested in the text.
\end{figure}

Finally, the trilayer and four layer devices show the expected tunneling behavior through an undoped semiconductor. The formation of a reduced quasiparticle gap (black arrows in Figure 3(g)) adds support to the hybridization model above. Quasiparticles tunnel from the reduced gap in the hybridized layer through a central undoped semiconducting layer (see Figure 3(k) for a simple model) which provides the tunnel barrier. This is reminiscent of earlier niobium junctions with a thin aluminum layer resulting in a reduced gap\cite{gurvitch1983high}. The four layer device (Figures 3(d, h, i)) follows this trend with a deeper quasiparticle gap. These simple models qualitatively explain the layer dependence of the presented junctions.

\section{Conclusion}
In conclusion, we have presented interlayer transport measurements on vertical MoRe-MoS$_2$-MoRe Josephson junctions. We found that the transport characteristics are dependent on the number of MoS$_{2}$ layers between the MoRe electrodes and, in particular, we observe a threshold between metallic-like weak link behavior and tunneling-like weak link behavior occurring between the bilayer and trilayer flake thicknesses. We propose that the metallic characteristics (appearance of MAR) of the single and bilayer devices are due to hybridization with the electrodes, as suggested in literature, that results in metalized MoS$_2$ layers. In the three and four layer devices the uncoupled layers (not directly in contact with the MoRe contacts) provide a tunnel barrier which reduces the critical current densities and results in more well-defined quasiparticle transport gaps. We anticipate the extension of this device architecture to other 2D materials and an interesting technological direction would be the use of insulating boron-nitride layers as uniform tunnel barriers replacing the standard AlOx barrier where less than 10\% of the barrier area is active in transport \cite{zeng2015}.    

\section{Acknowledgments}
The authors would like to thank Teun M. Klapwijk for insightful discussions. This work was supported by the Dutch organization for Fundamental Research on Matter (FOM), the Ministry of Education, Culture, and Science (OCW), and the Netherlands Organization for Scientific Research (NWO). A.C.-G. acknowledges financial support by the Fundacion BBVA through the fellowship “I Convocatoria de Ayudas Fundacion BBVA a Investigadores, Innovadores y Creadores Culturales” (Semiconductores Ultradelgados: hacia la optoelectronica flexible) and from the MINECO (Ramon y Cajal 2014 program, RYC- ́ 2014-01406) and from the MICINN (MAT2014-58399-JIN).

\section*{References}
\bibliography{Mos2_bib}

\include{Supplement}

\end{document}

%% file: Supplement.tex
\newpage
\title{Supporting Information:} \title{Thickness dependent interlayer transport in vertical MoS2 Josephson junctions}
\author{Joshua O. Island$^1$, Gary A. Steele$^1$, Herre S.J. van der Zant$^1$, Andres Castellanos-Gomez$^2$}
\address{$^1$ Kavli Institute of Nanoscience, Delft University of Technology, Lorentzweg 1, 2628 CJ Delft, The Netherlands.}
\address{$^2$Instituto Madrileno de Estudios Avanzados en Nanociencia (IMDEA Nanociencia), Campus de Cantoblanco, E-28049 Madrid, Spain}
\ead{j.o.island@tudelft.nl}

\newpage
\textbf{1. Photoluminescence (PL) and atomic force microscopy (AFM) determination of flake thickness}\\
	Figure S1 shows optical images, AFM profile scans, and PL spectra for devices having thicknesses of 1-4 MoS2 layers. Figure S21(a) shows three junctions fabricated using a flake having a single layer and bi layer region. The AFM line scans (take at the location of the dotted white lines in Figure S1(a)) are shown in Figure S1(c) and correspond to single and bilayer flakes. Additionally, Raman normalized PL spectra taken at the same approximate locations of the AFM scans in Figure S1(a) are shown in Figure S1(d). The monolayer flake has a much stronger PL spectra due to the direct band gap nature as reported in literature. Adding more layers results in a more indirect band gap and weaker PL response. Figure S1(b) shows two junctions fabricated using a flake with three and four layers. The corresponding AFM scans and PL spectra are shown in Figure S1(c) and S1(d) respectively.  
 
\textbf{2. Magnetic field dependence of the critical current}\\
	Figure S2 shows the magnetic field dependence for devices E, A and B, two single layer and one bilayer MoS2 junction. The field is applied parallel to the sandwich, alloying flux to penetrate the MoS2 flake. Ideally, such measurements could reveal information about the spatial uniformity of the critical current in the junctions. However, all three devices show a stochastic switching of the critical current as a function of magnetic field, which we attribute to motions of pinned vortices in the disordered superconducting MoRe leads at fields far above the first critical field (Hc1) of the type-II superconductor.

\textbf{3. Critical current densities and IcRn products as a function of layer number}\\
The critical current densities, like the critical currents, for each junction decrease with increasing layer number (see Figure S3(a)). The single and bilayer devices are orders of magnitude higher than the trilayer device indicating the crossover to tunneling suggested in the main text. The IcRn products for each device are shown in Figure S3(b) scaled by the superconducting gaps where we have used the bulk value for the single layer device and the proximity induced values for the 2, 3, and 4 layer devices ($\Delta=1.3$ meV, $1.1$ meV, $\Delta=0.75$ meV, $0.6$ meV for devices A-D respectively).  

\begin{figure}
\centerline{\includegraphics[width=5in]{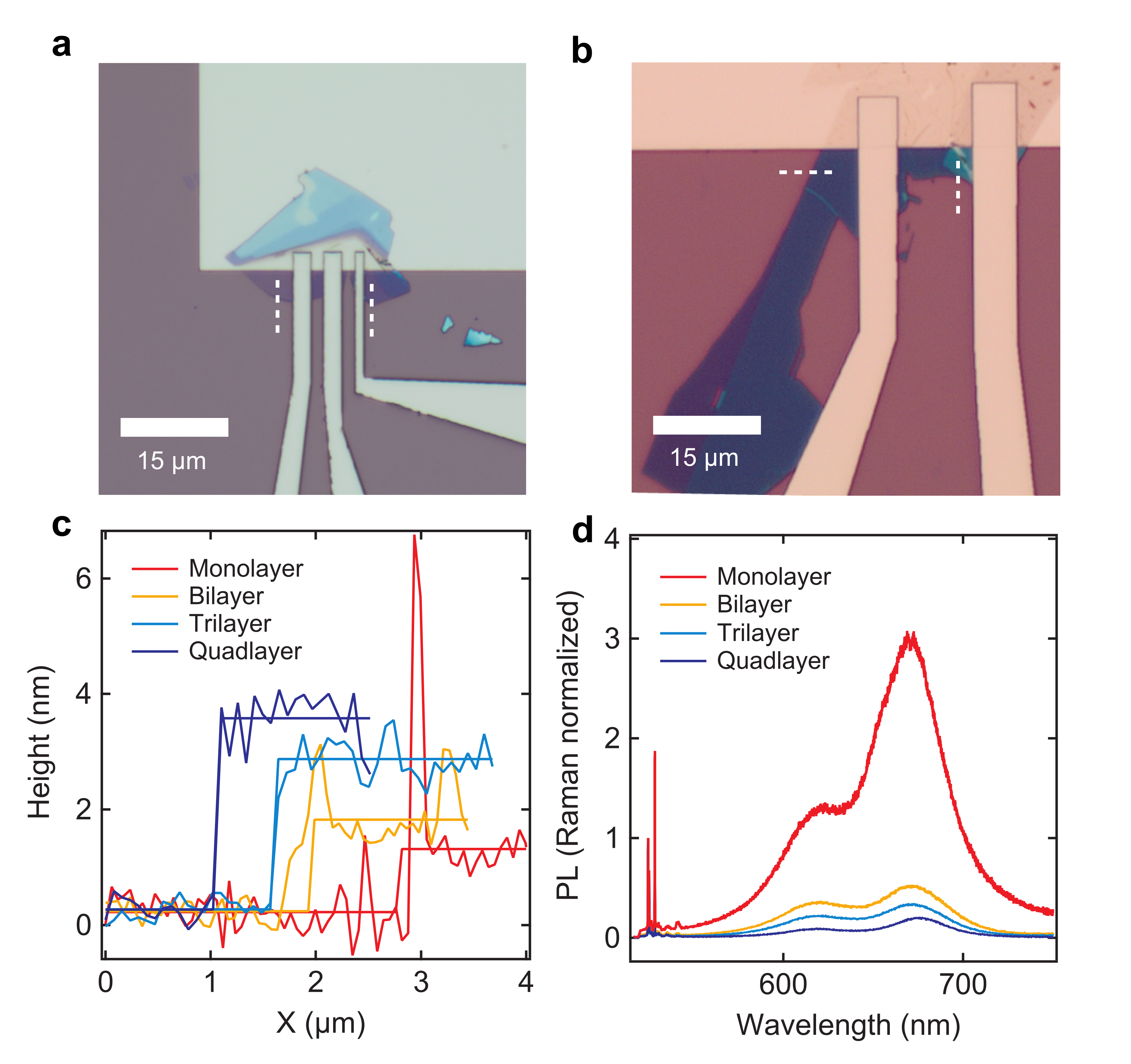}}
\textbf{Figure S1.} (a)  Optical image of three junctions made from an MoS2 flake having a single layer (left junction) and bilayer regions (right junction). (b) Optical image of two junctions fabricated with MoS2 flakes having three (left junction) and four layers (right junction). (c) AFM line profiles taken at the positions of the dotted lines in panels (a) and (b). (d) PL spectra recorded at the approximate locations of the AFM scans in panel (a) and (b).
\end{figure}

\begin{figure}
\centerline{\includegraphics[width=5in]{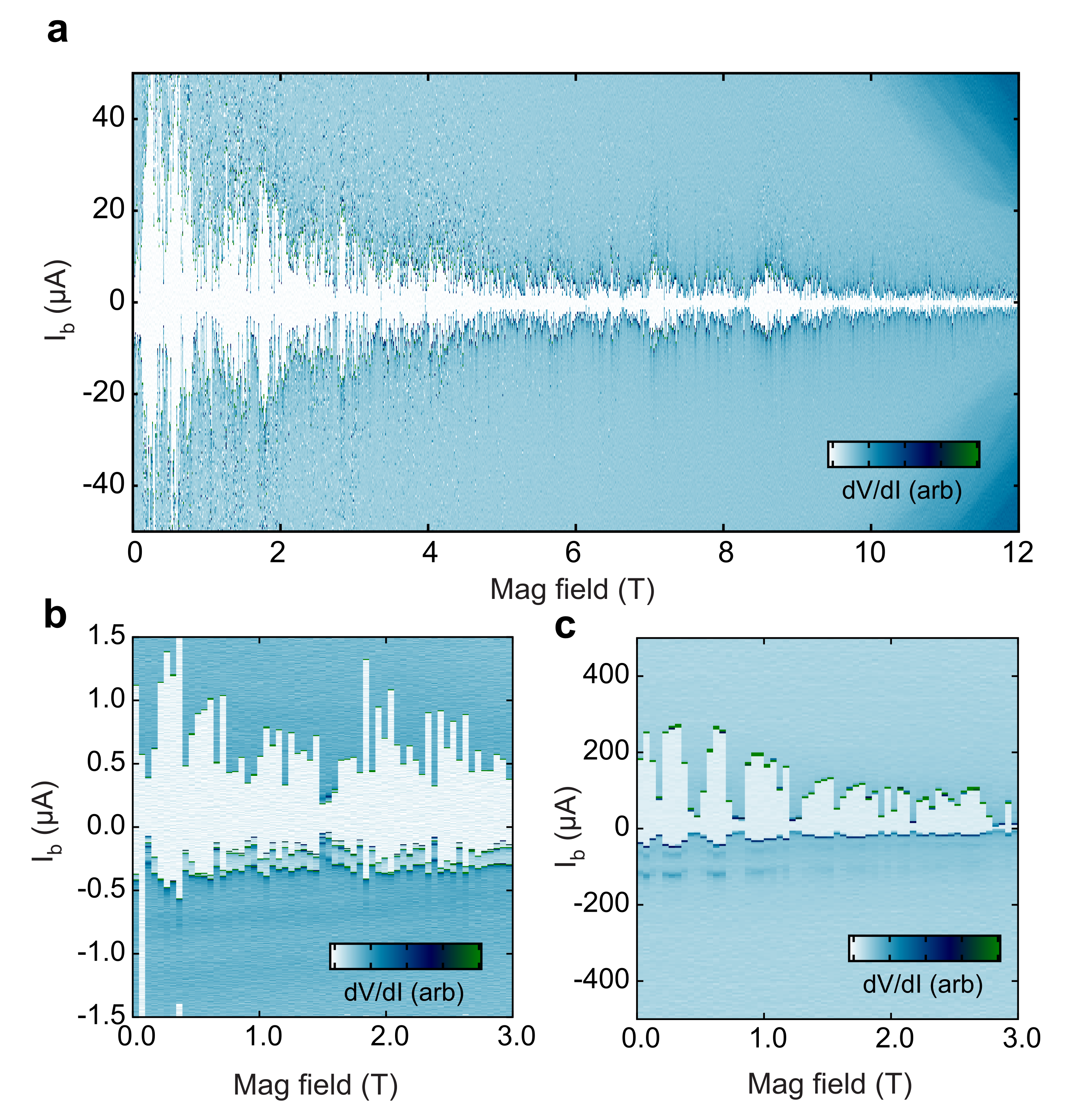}}
\textbf{Figure S2.} (a) Color plot of the magnetic field dependence of device E with field applied parallel to the junction sandwich. The differential conductance (dI/dV) is plotted as a function of current bias and magnetic field. (b) Same plot for device A. (c) Same plot for device B.
\end{figure}

\begin{figure}
\centerline{\includegraphics[width=6in]{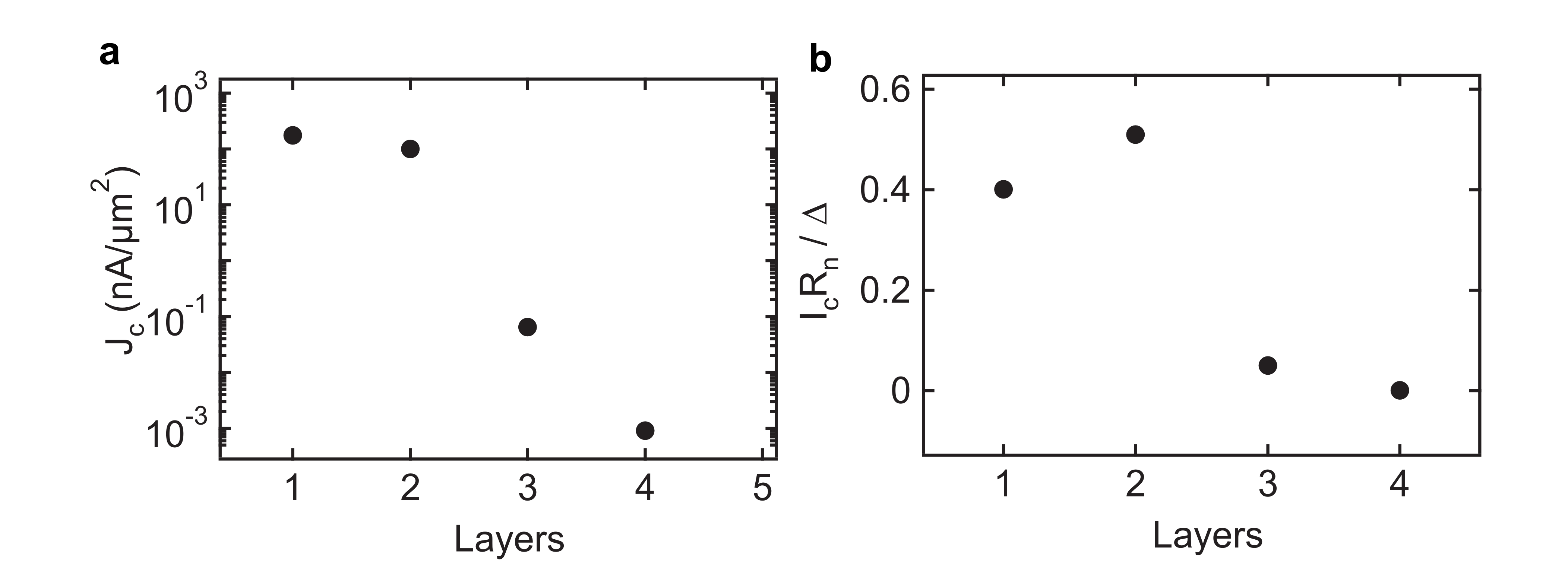}}
\textbf{Figure S3.} (a) Critical current densities for devices A-D having 1-4 layers respectively. (b) IcRn products normalized by the superconducting gap as a function of layer number for devices A-D.
\end{figure}